\documentclass[11pt,epsf,letterpaper]{article}%
\usepackage{color}
\usepackage{amsmath}
\usepackage{amsfonts}
\usepackage{verbatim}
\usepackage{amssymb}
\usepackage{graphicx}
\usepackage{epstopdf}
\usepackage{mathrsfs}%
\providecommand{\U}[1]{\protect\rule{.1in}{.1in}}
\begin{document}

\title{On attractor mechanism of $AdS_{4}$ black holes}
\author{$^{(1,2)}$Andr\'{e}s Anabal\'{o}n and $^{(3)}$Dumitru Astefanesei.\\\textit{$^{(1)}$Departamento de Ciencias, Facultad de Artes Liberales y}\\\textit{Facultad de Ingenier\'{\i}a y Ciencias, Universidad Adolfo
Ib\'{a}\~{n}ez,}\\\textit{Av. Padre Hurtado 750, Vi\~{n}a del Mar, Chile.}\\\textit{$^{(2)}$ Universit\'{e} de Lyon, Laboratoire de Physique, UMR 5672,
CNRS,}\\\textit{\'{E}cole Normale Sup\'{e}rieure de Lyon}\\\textit{46 all\'{e} d'Italie, F-69364 Lyon Cedex 07, France.}\\\textit{$^{(3)}$Instituto de F\'\i sica, Pontificia Universidad Cat\'olica de
Valpara\'\i so,} \\\textit{ Casilla 4059, Valpara\'{\i}so, Chile.}}
\maketitle

\begin{abstract}
We construct a general family of exact non-extremal $4$-dimensional black
holes in AdS gravity with $U(1)$ gauge fields non-minimally coupled to a
dilaton and a non-trivial dilaton potential. These black holes can have
spherical, toroidal, and hyperbolic horizon topologies. We use the entropy
function formalism to obtain the near horizon data in the extremal limit. Due
to the non-trivial self-interaction of the scalar field, the zero temperature
black holes can have a finite horizon area even if only the electric field is
turned on.

\end{abstract}

\section{Introduction.}

In recent years, it became clear that the Anti-deSitter(AdS)/conformal field
theory (CFT) may be also an useful tool for studying strongly-coupled toy
models of condensed matter systems (see, e.g., \cite{Horowitz:2010gk} and
references therein).

A careful analysis of charged dilaton black holes in AdS$_{4}$ was initiated
in \cite{Goldstein:2009cv}. The focus of that work was on extremal and
near-extremal solutions when the scalar potential is constant (proportional to
the cosmological constant). Although a lot of work was done in this direction,
it seems that \textit{exact} hairy AdS black hole solutions with a non-trivial
dilaton potential are not easy to construct. In the seminal work
\cite{Gibbons:1987ps} (see, also, \cite{ Dobiasch:1981vh}), a concrete method
for solving the equations of motion was proposed and families of
\textit{asymptotically flat} hairy black hole solutions were constructed. That
is, to rewrite the equations of motion as Toda equations and use the
well-known techniques to solving them. Unfortunately, this method is not
powerful enough to get general exact AdS$_{4}$ hairy black holes
\cite{Duff:1999gh}.

In this work we present a general family of \textit{exact} $4$-dimensional
non-extremal black holes in $AdS$ gravity with $U(1)$ gauge fields
non-minimally coupled to a scalar field. Recently, a class of exact $AdS_{4}$ black
hole solutions was obtained \cite{Lu:2013ura}, though in that work a different
method was used (see also \cite{Klemm:2012yg, Martinez:2006an, Xu:2013nia,
Bardoux:2012tr, Toldo:2012ec}). Our solutions are more general and are of
particular interest to clarifying aspects of the AdS/CMT correspondence
because they can provide information about distinct holographic phases of
matter \cite{Salvio:2012at}. Also, they can be useful in the context of 
fake supergravity \cite{Boonstra:1998mp}. We consider a non-trivial dilaton
potential and obtain the corresponding superpotential in a particular case.

However, there is another important motivation for our work. That is, these
solutions can exhibit new features which are absent for theories with a
trivial dilaton potential \cite{Goldstein:2009cv}. For example, the extremal
limit black hole solutions can have a finite horizon area even if just one
electric field is turned on. To the best of our knowledge these are the first
\textit{analytic} asymptotically AdS black holes with this property. From this
point of view, this is an interesting example where the attractor mechanism
\cite{Ferrara:1995ih} is at work.\footnote{It is well known that extremal
black holes exhibit the attractor mechanism regardless of
supersymmetry\cite{Goldstein:2005hq, Sen:2005wa} and this is at the basis of
computing the entropy of extremal non-BPS black hole \cite{Dabholkar:2006tb,
Astefanesei:2006sy}.} As was observed in \cite{Goldstein:2005hq,
Astefanesei:2007vh}, the moduli flow can be interpreted as an RG flow in the
context of AdS/CFT duality. Therefore, different kinds of
attractors will characterize the IR behavior of the dual CFT which is at zero
temperature but is now deformed by the addition of a chemical potential.

The paper is organized as follows: in Section $2$ we present the set-up. Then,
in Section $3$, we construct the most general solution for an arbitrary
exponential dilaton coupling of the gauge field and present some concrete
examples. In Section $4$ we discuss the extremal limit in the context of
attractor mechanism. We conclude the paper with comments and present some 
possible further directions in Section $5$.

\section{General formalism}

In this section we follow closely \cite{Anabalon:2013qua} (see, also,
\cite{Anabalon:2012ta}) where the asymptotically flat `cousins' of our black
hole solutions were constructed.

We are interested in a generic action of the form
\begin{equation}
\label{I}I[g_{\mu\nu},A_{\mu},\phi]=\frac{1}{2\kappa}\int d^{4}x\sqrt
{-g}\left[  R-\frac{1}{4}e^{\gamma\phi}F^{2}-\frac{1}{2}\partial_{\mu}%
\phi\partial^{\mu}\phi-V(\phi)\right]
\end{equation}
where the gauge coupling and potential are functions of the dilaton and we use
the convention $\kappa=8\pi G_{N}$. Since we set $c=1=\hbar$, $\left[
\kappa\right]  = M_{P}^{-2}$ where $M_{P}$ is the reduced Planck mass. The
equations of motion for the gauge field, dilaton, and metric are
\[
\nabla_{\mu}\left(  e^{\gamma\phi}F^{\mu\nu}\right)  =0
\]

\[
\frac{1}{\sqrt{-g}}\partial_{\mu}\left(  \sqrt{-g}g^{\mu\nu}\partial_{\nu}%
\phi\right)  -\frac{\partial V}{\partial\phi}-\frac{1}{4}\gamma e^{\gamma\phi
}F^{2}=0
\]

\[
R_{\mu\nu}-\frac{1}{2}g_{\mu\nu}R=\frac{1}{2}\left[  T_{\mu\nu}^{\phi}%
+T_{\mu\nu}^{EM}\right]
\]
where the stress tensors of the matter fields are%

\[
T_{\mu\nu}^{\phi}=\partial_{\mu}\phi\partial_{\nu}\phi-g_{\mu\nu}\left[
\frac{1}{2}\left(  \partial\phi\right)  ^{2}+V(\phi)\right]
\,\,\,\,\,\,\,\,\,T_{\mu\nu}^{EM}=e^{\gamma\phi}\left(  F_{\mu\alpha}F_{\nu
}^{\cdot\alpha}-\frac{1}{4}g_{\mu\nu}F^{2}\right)
\]

The ansatz for the metric is more general than in asymptotically flat case
\cite{Anabalon:2013qua}, because in AdS there also are black holes with
toroidal and hyperbolic horizon topologies:
\begin{equation}
ds^{2}=\Omega(x)\left[  -f(x)dt^{2}+\frac{\eta^{2}dx^{2}}{f(x)}+d\Sigma
_{k}\right]  \label{Ansatz}%
\end{equation}
where the parameter $\eta$ was introduced to obtain a dimensionless radial
coordinate $x$ and $\Omega(x)$ is the conformal factor. $d\Sigma_{k}$ is a
surface of constant curvature normalized to be $k=\pm1$ or $0$ and which can
be conveniently parameterized as\footnote{The usual canonical form is obtained
with the following change of coordinates: $k =1\Longrightarrow y=\cos\theta$
and $k =-1\Longrightarrow y=\sinh\theta$.}%

\begin{equation}
d\Sigma_{k}=\frac{dy^{2}}{1-ky^{2}}+\left(  1-ky^{2}\right)  d\varphi^{2}%
\end{equation}

This is the most general static asymptotically locally AdS ansatz,
characterized by only two unknown functions. The equation of motion for the
gauge field and Bianchi identity can be solved by the following ansatz of the
field strength:
\begin{equation}
F=Qe^{-\gamma\phi}dx\wedge dt+P{dy}\wedge d\varphi\label{GF}%
\end{equation}

The strategy for finding exact solutions is similar with the one in the
asymptotically flat case \cite{Anabalon:2013qua} and we do not repeat all the
steps here. We use the same conformal factor%

\begin{equation}
\Omega(x)=\frac{\nu^{2}x^{\nu-1}}{\eta^{2}\left(  x^{\nu}-1\right)  ^{2}}
\label{CF}%
\end{equation}

In these coordinates, the scalar can be written in a nice form
\begin{equation}
\phi(x)=l_{\nu}^{-1}\ln(x)+\phi_{0} \label{scalar 2}%
\end{equation}
where $l_{\nu}=\left(  \nu^{2}-1\right)  ^{-\frac{1}{2}}$ plays the role of a
characteristic length scale of the dilaton. There is a $\pm$ ambiguity in the
integration of the dilaton equation, which corresponds to a discrete
degeneration in the black hole family. Indeed, from (\ref{CF}) it follows that
the conformal factor has a pole of order two at $x=1$ where the conformal
infinity is `located'. The fact that (\ref{CF}) is regular in the region
$x\in\left(  0,1\right)  $ and $x\in\left(  1,\infty\right)  $ allows to pick
any of these intervals as the domain of coordinates, one corresponding to a
negative scalar and the other corresponding to a positive one.

The remaining metric function satisfies the following differential equation:
\begin{equation}
\frac{1}{\Omega}\left(  \Omega f^{\prime}\right)  ^{\prime}+2\eta^{2}%
k-\frac{e^{-\gamma\phi}Q^{2}+e^{\gamma\phi}\eta^{2}P^{2}}{\Omega}=0
\label{FE1}%
\end{equation}
which can be exactly integrated.

\section{Exact non-extremal solutions.}

In this section, we construct static asymptotically flat non-extremal black
holes for a model with one scalar field (dilaton) non-minimally coupled to a
gauge field and a non-trivial dilaton potential. We consider an arbitrary
exponential dilaton coupling of the gauge field. We obtain a class of
solutions that generalize the Gibbons-Maeda solutions \cite{Gibbons:1987ps} in AdS.

\subsection{Generalized Gibbons-Maeda solution in AdS.}

Gibbons and Maeda \cite{Gibbons:1987ps} found the black hole of the
Einstein-Maxwell-dilaton theory defined by the action
\begin{equation}
I[g_{\mu\nu},A_{\mu},\phi]=\int d^{4}x\sqrt{-g}\left(  R-\frac{1}{4}%
e^{\gamma\phi}F^{2}-\frac{1}{2}\left(  \partial\phi\right)  ^{2}\right)
\end{equation}
which matches the action (\ref{I}) when the dilaton potential vanishes.

First step in our analysis is to rewrite the solutions of
\cite{Gibbons:1987ps} in the coordinates similar with the ones we use. This is
useful to make the contact with the old known solutions and also helps to gain
some intuition to understanding the general class of solutions we will present.

For simplicity, we set $\gamma=\left(  \frac{\nu+1}{\nu-1}\right)  ^{1/2}$,
then the Gibbons-Maeda solution is
\begin{equation}
ds^{2}=\Omega(x)\left(  -f(x)dt^{2}+\frac{\eta^{2}dx^{2}}{f(x)}+d\Sigma
_{1}\right)
\end{equation}%
\begin{equation}
F(x)=\frac{\eta^{2}x^{2}(x^{\nu}-1)^{2}k}{x^{\nu}\nu^{2}}-\frac{\eta^{2}%
x^{2}(x^{\nu}-1)^{3}Q^{2}}{2\nu^{3}(\nu-1)x^{2\nu}}\label{GM}%
\end{equation}
with the gauge field and scalar
\begin{equation}
F=\frac{Q}{x^{\nu+1}}dx\wedge dt\qquad\phi=l_{\nu}^{-1}\ln(x)\qquad
\end{equation}
The functions $\Omega(x)$ and $l_{\nu}$ are given in the previous section.
Since the Gibbons-Maeda solution is asymptotically flat, $k$ is fixed to $1$
so that the horizon topology is spherical. With this parameterization, it is
possible to consider positive or negative $\gamma$ depending if one consider
$x>1$ or $x<1$. In both cases, $x=1$ defines the asymptotic region of the
spacetime as can be seen by the pole of order two.

We would like to point out that the new parameter $\nu$ labels different
solutions, for example $\nu=-1$, $\gamma=0$, corresponds to
Reissner-Nordstr{\"{o}}m black hole and for $\nu>1$ we obtain `hairy'
solutions. Therefore, the Gibbons-Maeda solution is continuously connected
with the Reissner-Nordstr\"{o}m solution when $\nu=-1$. Indeed, in this case
$\gamma=0$, $\phi=0$, and the change of coordinates \thinspace%
\begin{equation}
x=1+\frac{1}{\eta\rho}%
\end{equation}
brings the solution to the well-known form
\begin{equation}
ds^{2}=-h(\rho)dt^{2}+\frac{d\rho^{2}}{h(\rho)}+\rho^{2}d\Sigma_{1}%
\end{equation}
where $h(\rho)=k-\frac{2M}{\rho}+\frac{q^{2}}{\rho^{2}}$ and the parameters of
the two solutions are related by $\eta=-\frac{1}{2M}\left(  k+\frac{Q^{2}}%
{4}\right)  $ and $q=\frac{Q}{2\eta}$.

What is interesting is that even in the presence of a non-trivial dilaton
potential, we can still obtain asymptotically flat solutions
\cite{Anabalon:2013qua}. We also present this step because it is useful for
generating more general $AdS$ solutions. We consider the following non-trivial
self-interaction of the scalar field
\begin{equation}
V_{\alpha}(\phi) =2\alpha\left[  \frac{\nu-1}{\nu+2}\sinh(\phi l_{\nu}\left(
\nu+1\right)  )-\frac{\nu+1}{\nu-2}\sinh(\phi l_{\nu}\left(  \nu-1\right)
)+4\frac{\nu^{2}-1}{\nu^{2}-4}\sinh(\phi l_{\nu})\right]
\end{equation}
for which there still exists an asymptotically flat black hole solution for
$k=1$:
\begin{equation}
f_{\alpha}(x) =\frac{\eta^{2}x^{2}(x^{\nu}-1)^{2}k}{x^{\nu}\nu^{2}}-\frac
{\eta^{2}x^{2}(x^{\nu}-1)^{3}Q^{2}}{2\nu^{3}(\nu-1)x^{2\nu}}+\alpha\left[
\frac{\nu^{2}}{\nu^{2}-4}+(-x^{2}+\frac{x^{\nu+2}}{\nu+2}-\frac{x^{2-\nu}}%
{\nu-2})\right]  \label{A1}%
\end{equation}
Here we keep the general expression with $k$ arbitrarily because that is what
we need to write down the asymptotically AdS solutions. In flat space, we
should fix $k=1$, which corresponds to the spherical horizon topology. It is
clear that the Gibbons-Maeda solution can be obtained for $\alpha=0$. The role
of the subindex $\alpha$ in the scalar field potential and metric function is
only to simplify the notation in what follows.

Now we are ready for the last step. That is, we generalize the solution
(\ref{A1}) in the presence of a negative cosmological constant. After lengthy
computations we obtain:
\begin{equation}
f_{\alpha,\Lambda}(x)=f_{\alpha}-\frac{\Lambda}{3}%
\end{equation}
and the dilaton potential is
\begin{equation}
V_{\alpha,\Lambda}(\phi)=V_{\alpha}(\phi)+\frac{\Lambda\left(  \nu
^{2}-4\right)  }{3\nu^{2}}\left[  \frac{\nu-1}{\nu+2}\exp(-\phi l_{\nu}\left(
\nu+1\right)  )+\frac{\nu+1}{\nu-2}\exp(\phi l_{\nu}\left(  \nu-1\right)
)+4\frac{\nu^{2}-1}{\nu^{2}-4}\exp(-\phi l_{\nu})\right]  \label{Pot}%
\end{equation}

One can check that the metric is asymptotically AdS by computing the Ricci
scalar that is at the boundary $\left.  R\right\vert _{x=1}=4\Lambda$. Also,
one can easily check that at the boundary $x=1$ where the dilaton vanishes the
dilaton potential is $V(0)=2\Lambda$.

Furthermore, these solutions can be generalized for an electromagnetic field
\begin{equation}
F=\frac{Q}{x^{\nu+1}}dx\wedge dt+Pdy\wedge d\varphi
\end{equation}
with the following metric function and dilaton potential%

\begin{equation}
f_{\alpha,\lambda,P}(x)=f_{\alpha,\lambda}(x)+\frac{P^{2}\eta^{4}x^{4}}%
{\nu^{2}(\nu+2)}\left(  \frac{1}{4}-\frac{2x^{\nu}}{\nu+4}+\frac{x^{2\nu}%
}{2\left(  \nu+2\right)  }\right)
\end{equation}

\begin{align}
V_{\alpha,\lambda,P}(\phi)  &  =V_{\alpha,\lambda}(\phi)+\frac{P^{2}\eta
^{4}x^{3}\left(  \nu+1\right)  }{\nu^{4}\left(  \nu+2\right)  \left(
\nu+4\right)  }\left(  -\frac{(\nu+4)e^{-\nu l_{\nu}\phi}}{2\left(
\nu+1\right)  }+\frac{2(\nu+8)e^{-2\nu l_{\nu}\phi}}{\left(  \nu+2\right)
}\right. \nonumber\\
&  \left.  -\frac{\nu^{3}-10\nu^{2}+48\nu+96}{4\left(  2+\nu\right)  }e^{-3\nu
l_{\nu}\phi}+\left(  -\nu^{2}+3\nu+8\right)  e^{-4\nu l_{\nu}\phi}-\frac
{(\nu+4)\left(  \nu+2\right)  }{4}e^{-5\nu l_{\nu}\phi}\right)
\end{align}
where $\lambda=\Lambda+\frac{3\eta^{2}P^{2}}{4(\nu+4)\left(  \nu+2\right)
^{2}}$.

\subsection{Examples}

We present examples when $\gamma=\sqrt{3}$ and $\gamma=1$ for which the
potential simplifies drastically. We consider only the branches for which the
dilaton is positive and so $x > 1$. Interestingly enough, these solutions are
smoothly connected with some solutions that can be embedded in string theory.
We are able to recover some well-known solutions \cite{Duff:1999gh,
Lu:2013ura} in some special limits and obtain a new dyonic solution in the
$\gamma=1$ case that can be embedded in string theory.

\subsubsection{$\gamma=\sqrt{3}$, $\nu=2$}

For this value of the coupling, we are able to obtain only $P=0$ case. The
dyonic solution is technically difficult to obtain because the dilaton
potential is complicated when $P$ is turned on.

The metric is again characterized by the same conformal factor $\Omega(x)$ and
the other metric function is
\begin{equation}
f(x)=-\frac{\Lambda}{3}-\frac{\eta^{2}Q^{2}\left(  x^{2}-1\right)  ^{3}%
}{16x^{2}}+\alpha(-\frac{x^{4}}{12}+\frac{x^{2}}{3}-\frac{1}{4}-\frac{1}{3}%
\ln(x))+\frac{k\eta^{2}\left(  x^{2}-1\right)  ^{2}}{4}%
\end{equation}

The dilaton potential simplifies in this case and becomes
\begin{equation}
V(\phi)=2\Lambda\cosh(\phi l_{2})+\frac{\alpha}{2}\left[  4\phi l_{2}%
\cosh(\phi l_{2})-3\sinh(\phi l_{2})-\frac{1}{3}\sinh(3\phi l_{2})\right]
\end{equation}

When $\Lambda=\alpha=0$ we obtain the electrically charged KK black hole that
can be embedded in $N=2$ SUGRA. In flat space the extremal limit and BPS limit
coincide and a naked singularity is obtained. When only $\alpha=0$ and
$k=\{0,1\}$ we obtain the AdS solutions discussed in detail in
\cite{Lu:2013ura} but in a different coordinate system, though we also have
the exact solution for $k=-1$. The non-extremal solution for $k=1$ is a
regular black hole, but a domain wall for $k=\{0, -1\}$ when $\alpha=0$. In
AdS the extremal limit and BPS limit do not coincide. However, we are going to
present an argument in the next section that shows that the corresponding zero
temperature limit is still a naked singularity.

Interestingly enough, once we consider the $\alpha$-part in the potential, we
obtain regular non-extremal black hole solutions with spherical, toroidal, and
hyperbolic horizon topologies corresponding to $k=\{-1, 0, 1\}$. In this case,
in the extremal limit we obtain regular zero temperature regular black holes
with a finite horizon area.

\subsubsection{$\gamma=1$}

This limit is more subtle and it was explained for asymptotically flat black
holes in \cite{Anabalon:2013qua} and we do not want to present the details here.

In this case the metric is (the form of the metric is slightly different than
the general ansatz (\ref{Ansatz}) in Section $2$):%

\begin{equation}
ds^{2}=\Omega(x)\left[  -f(x)dt^{2}+\frac{\eta^{2}dx^{2}}{x^{2}f(x)}%
+d\Sigma_{k}\right]
\end{equation}
where
\begin{equation}
f(x)=-\frac{\Lambda}{3}+\alpha\left[  \frac{(x^{2}-1)}{2x}-\ln(x)\right]
+k\eta^{2}\frac{(x-1)^{2}}{x}+\frac{\eta^{2}}{2x^{2}}\left(  x-1\right)
^{3}\left(  P^{2}\eta^{2}x-Q^{2}\right)
\end{equation}

The gauge field and dilaton potential are
\begin{equation}
F=\frac{Q}{x^{2}}dx\wedge dt+Pdy\wedge d\varphi
\end{equation}%
\begin{equation}
V(\phi)=\left(  \frac{\Lambda}{3}+\alpha\phi\right)  \left(  4+2\cosh
(\phi)\right)  -6\alpha\sinh(\phi) \label{Pot1}%
\end{equation}

When $\Lambda=\alpha=0$ the solution is asymptotically flat and can be
embedded in $N=4$ SUGRA. When only $\alpha=0$ the situation is similar with
the one in $\gamma=\sqrt{3}$ case and the black hole/domain wall solutions can
be embedded in string theory. The extremal limit when $\alpha\neq0$ is
discussed in detail in the next section.

\section{Extremal limit and attractor mechanism}

As we have already mentioned, when $\alpha$ is turned on, we are going to show
tat the extremal limit is a regular zero-temperature black hole with finite
horizon area even when $P=0$. This is possible due to the existence of a
non-trivial dilaton potential. In this case, there is a modified effective
potential \cite{Goldstein:2005hq}:
\begin{equation}
V_{eff}(\phi)=\frac{1}{b^{2}}\left[  e^{-\phi}Q^{2} + e^{\phi}P^{2} \right]  +
2b^{2} V(\phi)
\end{equation}
that controls the dilaton flow.

When $P=0$ and $V(\phi)=constant$ this effective potential can not have a
minimum and so there no exist regular extremal black hole with finite horizon
area.\footnote{When a second charge is turned on, the effective potential has
a minimum and regular solutions exist.} When the dilaton potential is
non-trivial $V(\phi)\neq0$, it could be possible that the effective potential
has a minimum at the horizon and in the extremal limit one can obtain a black
hole with finite entropy.

Since the effective potential method and entropy function formalism are
equivalent in the near horizon limit,\footnote{Since we have exact solutions,
we could find the near horizon directly for each solution. However, such a
computation is difficult and, particularly, not very illuminating.} we prefer
to use the entropy function of Sen \cite{Sen:2005wa}\footnote{The entropy function 
formalism for AdS black holes was initiated in \cite{Morales:2006gm} and a detailed 
analysis of entropy function and near horizon data for AdS black hole
can be found in \cite{Astefanesei:2010dk}.} to prove the existence of
regular extremal black holes with finite horizon area when $P=0$.

The near horizon geometry of $4-$dimensional static extremal black holes has
an enhanced symmetry of $AdS_{2}\times S^{2}$, and unlike the non-extremal
case, it is a solution of the equations of motion. We consider the following
ansatz for the metric and gauge field:
\begin{equation}
ds^{2}=v_{1}(-r^{2}dt^{2}+\frac{dr^{2}}{r^{2}})+v_{2}\left[  \frac{dy^{2}%
}{1-y^{2}}+(1-y^{2})d\phi^{2}\right]  \qquad F=Edx\wedge dt
\end{equation}

The entropy function is related to the action evaluated at the horizon by:
\begin{equation}
S(E,v_{1},v_{2},u)=2\pi(Eq-I_{H})=2\pi\lbrack Eq-(-2v_{2}+2v_{1}-v_{1}%
v_{2}V(u)+2^{-1}v_{1}^{-1}v_{2}e^{u}E^{2})]
\end{equation}
where $q$ is the physical charge and $u$ is the value of the scalar field at
the horizon. The attractor equations are
\begin{align}
\frac{\partial S}{\partial E} &  =q-\frac{v_{2}}{v_{1}}e^{u}E=0\label{Eq1}\\
\frac{\partial S}{\partial v_{1}} &  =2-v_{2}V(u)-2^{-1}v_{1}^{-2}v_{2}%
e^{u}E^{2}=0\label{Eq2}\\
\frac{\partial S}{\partial v_{2}} &  =-2-v_{1}V(u)+2^{-1}v_{1}^{-1}e^{u}%
E^{2}=0\label{Eq3}\\
\frac{\partial S}{\partial u} &  =-v_{1}v_{2}V^{\prime}(u)+2^{-1}v_{1}%
^{-1}v_{2}e^{u}E^{2}=0\label{Eq4}%
\end{align}

The equation (\ref{Eq1}) can be easily solved and then find $v_{1}$ and $v_{2}$ as
functions of the physical charge and the value of the dilaton at the horizon.
A further manipulation of the equations allows to implicitly find the value of
the scalar at the horizon for $q^{2}\neq0$
\begin{equation}
q^{2}\left(  V^{\prime}+V\right)  ^{2}=8e^{u}V^{\prime} \label{Atractor}%
\end{equation}

Let us consider now (\ref{Pot1}) and replace it in (\ref{Atractor}) --- for
$\alpha=0$ we get
\begin{equation}
\frac{q^{2}}{l^{2}}=-\frac{4e^{u}\sinh u}{\left(  \sinh u+\cosh u+2\right)
^{2}}%
\end{equation}
which has real solutions only when $u<0$, which corresponds to the rank of
coordinates $x<1$. Since we are interested in the $x>1$ branch we conclude, as
expected, that the extremal limit is a naked singularity. However, when
$\alpha\neq0$ the equation (\ref{Atractor}) yields%

\begin{equation}
\frac{q^{2}}{l^{2}}=\frac{4e^{u}\left(  2\alpha l^{2}+\alpha l^{2}u\sinh
u-2\alpha l^{2}\cosh u-\sinh u\right)  }{\left[  2+\sinh u+\cosh u+\alpha
l^{2}(2\cosh u-u\cosh u-u\sinh u-2u+3\sinh u-2)\right]  ^{2}} \label{at}%
\end{equation}

The equation (\ref{at}) is a quadratic equation for $\alpha$. It turns out
that the discriminant of this equation is always positive, which indicates the
existence of real solutions for any values of $q$, $l$, and $u$.

\section{Discussion and future directions}

In this paper, we have constructed a general class of four-dimensional AdS
dyonic black holes that generalize the black hole solutions of Gibbons and
Maeda in flat space \cite{Gibbons:1987ps}. The dilaton potential has two
parts, one that is controlled by the cosmological constant and the other one
controlled by an arbitrary parameter $\alpha$. Obviously, since the solutions
are asymptotically AdS, the part controlled by $\alpha$ does not play any role
at the boundary but controls the IR regime of the dual field theory. This is
why we find well defined moduli flows for the corresponding extremal black
holes even if just one gauge field (with only one charge) is turned on. Since
the extremal (attractor) horizons are infinitely far away in the bulk they do
not get distorted by changing the scalars boundary values and so, from this
point of view, the attractor mechanism acts as a no-hair theorem for extremal
black holes \cite{Astefanesei:2007vh}.

One can speculate that if the quantum corrections (which could be similar with
the $\alpha$-part of the potential) become relevant and the dilaton potential
gets corrected, naked singularities can get dressed with a horizon and in this
case the IR point of the RG flow is well defined. Sure, it is not clear if our
solutions can be embedded in a fundamental theory like string theory for
$\alpha\neq0$, but the main point is that these are generic features and a
similar situation may exist for `stringy' solutions.

When $\alpha=0$ we were able to recover some known solutions that can be
embedded in string theory \cite{Duff:1999gh, Lu:2013ura}. In this particular
case, we can explicitly obtain the superpotential associated with the dilaton
potential. The solution becomes a domain wall in the planar case $k=0$, but in
the spherical case $k=1$ we can still obtain regular black hole solutions. The
superpotential equation is%
\begin{equation}
V(\phi)=2\left(  \frac{dW(\phi)}{d\phi}\right)  ^{2}-\frac{3}{2}W(\phi)^{2}%
\end{equation}
and when $\alpha=0$ we obtain%

\begin{equation}
W(\phi)=l^{-1}\left[  \frac{\left(  \nu+1\right)  }{\nu}\exp(\left(  \frac
{\nu-1}{2}\right)  \phi l_{\nu})+\frac{\left(  \nu-1\right)  }{\nu}%
\exp(-\left(  \frac{\nu+1}{2}\right)  \phi l_{\nu})\right]
\end{equation}
where it was set $\Lambda=-\frac{3}{l^{2}}$.

Let us end up this section by presenting a few future directions. Since we
have the superpotential, we can construct the corresponding RG flow as in
\cite{de Boer:1999xf}. Also an analysis similar with the one of
\cite{Elvang:2007ba} in the context of fake supergravity can be done. A
generalization of these solutions to $5$-dimensions is possible and we have
some preliminary results \cite{noi}. Our solutions can also be useful in the
context of the recent work \cite{Prester:2013fia}.

In \cite{Lu:2013ura} an intriguing thermodynamic feature of the gauged dyonic
black hole was pointed out. That is, the usual first-law of thermodynamics
does hold just in some special cases. However, in general, a new pair of
thermodynamic conjugate variables should appear in the first law. They
describe the potential and charge of the scalar hair that breaks some of the
asymptotic AdS symmetries. A similar analysis can be also done for our solutions.

\section{Acknowledgments.}

We would like to thank Andres Acena, David Choque, Robert Mann, and Henning
Samtleben for interesting discussions. A.A. would like to thank the laboratory
of physics at ENS, Lyon where this project was completed. Research of A.A. is
supported in part by the Fondecyt Grant 11121187 and by the CNRS project
``Solutions exactes en pr\'{e}sence de champ scalaire''. The work of D.A. is
supported by the Fondecyt Grant 1120446. DA would also like to thank Albert
Einstein Inst., Potsdam for warm hospitality during the course of this work.

\end{document}